\documentclass[%
 aip,
rsi,
 amsmath,amssymb,
preprint,%
]{revtex4-1}

\usepackage{graphicx}
\usepackage{dcolumn}
\usepackage{bm}

\usepackage[utf8]{inputenc}
\usepackage[T1]{fontenc}
\usepackage{mathptmx}
\usepackage{etoolbox}

\makeatletter
\def\@email#1#2{%
 \endgroup
 \patchcmd{\titleblock@produce}
  {\frontmatter@RRAPformat}
  {\frontmatter@RRAPformat{\produce@RRAP{*#1\href{mailto:#2}{#2}}}\frontmatter@RRAPformat}
  {}{}
}%
\makeatother
\begin{document}

\preprint{AIP/123-QED}

\title[Photothermal resistivity alignment of optical fibers to SNSPD]{Photothermal resistivity alignment of optical fibers to SNSPD}
\author{M. Bar\'{a}nek}
\email{martin.baranek@fmph.uniba.sk}

\author{D. Lorenc}%
\affiliation{ 
International Laser Centre, Ilkovi\v{c}ova 3, 84104, Bratislava, Slovakia}%
\affiliation{ 
RCQI, Slovak Academy of Sciences, D\'{u}bravsk\'{a} cesta 9, 84511, Bratislava, Slovakia}%

\author{T. \v{S}\v{c}epka}

\author{J. \v{S}olt\'{y}s}

\author{I. Vetrova} 

\author{\v{S}. Ha\v{s}\v{c}\'{i}k}
\affiliation{Institute of Electrical Engineering, Slovak Academy of Sciences, D\'{u}bravsk\'{a} cesta 9, 84511, Bratislava, Slovakia}%

\author{M. Grajcar}%
\author{P. Neilinger}%
\affiliation{ 
Department of Experimental Physics, Comenius University, 84248, Bratislava, Slovakia
}%

\date{\today}

\begin{abstract}
We demonstrate a straightforward optoelectronic fiber alignment technique for superconducting nanowire single-photon detectors (SNSPDs) that exploits the temperature-dependent resistance of the nanowire under optical absorption. The target nanowire is illuminated via the fiber, and the local absorption of light heats the wire, causing a change in its resistivity. Scanning the fiber over the nanowire, the change in its resistivity is monitored by lock-in amplifier, mapping the spatial photothermal response correlated to absorption and coupling efficiency. The maximum of the response corresponds to optimal fiber-SNSPD alignment. This method allows for aligning the fiber to the center of the meander with sub-micron precision.
The response is robust to variations in the angle and height of the fiber, providing an alternative or complement to fiber-to-chip alignment methods based on the back-reflection or transmission measurement.
\end{abstract}

\maketitle

\section{Introduction}\label{sec1}
Superconducting nanowire single-photon detectors\cite{Goltsman01} (SNSPDs) combine near-unity detection efficiency,\cite{Chang21} mHz~level dark count rate,\cite{Mueller21} sub-50~ps timing jitter\cite{Najafi15} and GHz-scale count rate, making them the detector of choice for quantum optics, deep-space communications\cite{Wollman24} and quantum key distribution.\cite{Chen2021,Ribezzo23}
The key element of these detectors is the current-biased superconducting nanowire with a width below 100~nm. The absorption of a photon is sufficient to disrupt the superconductivity, resulting in a voltage pulse. The superconductor, from which the nanowires are fabricated, is strongly disordered and, therefore, is characterized by high sheet resistance in the normal state (up to 1~k$\Omega/\square$),\cite{Zolotov23} which usually decreases with increasing temperature. 
Although intrinsic device performance is governed by material properties, nanowire geometry, and cryogenic biasing conditions, the overall system detection efficiency is constrained by the optical interface required to guide infrared light into micrometer-scale active areas.

To ensure efficient coupling of the light to the nanowire, various alignment methods are utilized,\cite{Dauler14review} often relying on wafer etching methods. Notably, in Refs. \onlinecite{Chang21,Miller11}, a keyhole-shaped window is etched into the substrate by the Bosch process. A zirconium sleeve fits with micrometer precision to the hole, to which the fiber ferrule is inserted. The fiber ferrule is then pushed inside the sleeve up to the point of contact with the SNSPD. This method allows the precise alignment of the fiber and the nanowire with fiber height above the nanowire of less than 10~$\mu$m.
However, if the aforementioned mechanical self-alignmnent is not possible, the back-reflection to the fiber,\cite{Erotokritou2018} or through-wafer transmission\cite{Orgiazzi09} is utilized for SNSPD alignment. 

In this work, we present an alternative alignment method grounded in the fundamental principle of SNSPD, specifically the absorption of infrared radiation and the consequent heating of the nanowire. The optical fiber is scanned above the substrate, and by measuring the resistance of the nanowire, the modulated laser beam periodically heats the nanowire (by $\approx$~0.1~K), resulting in a small change (in order of $\approx10^{-5}$) of the resistance. The AC response is then separated by AC coupling and then measured by a lock-in amplifier. The experimental setup is similar to the Lock-in thermography,\cite{Breitenstein03} although with the roles of input and output exchanged.

\begin{figure}[h]
\centering 
\includegraphics[width=\linewidth]{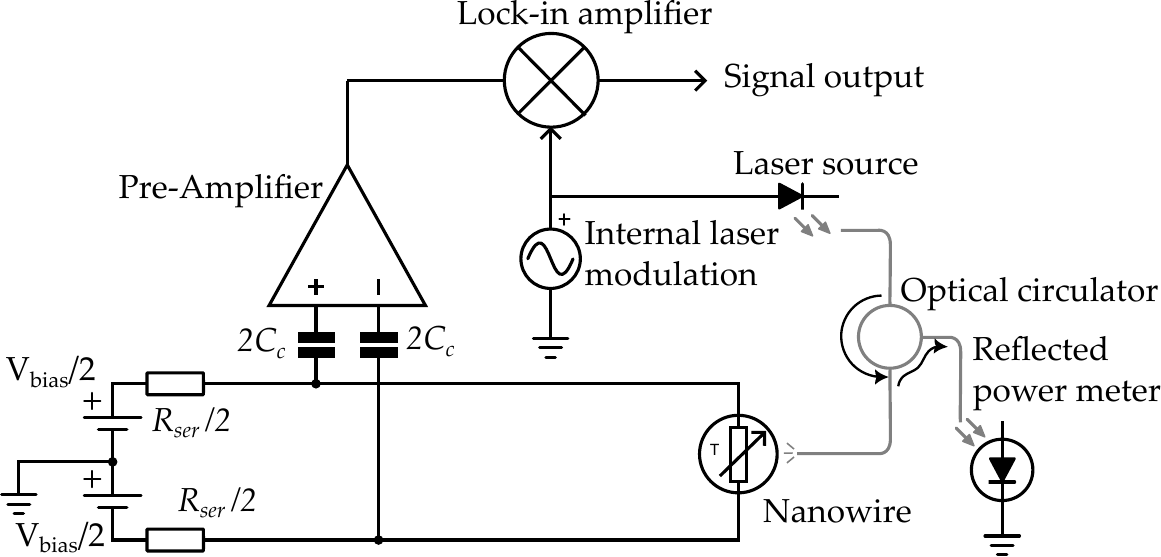}%
\caption{Scheme of experimental apparatus for photothermal resistivity alignment and optical reflectivity alignment. Optical fibers are grey-colored. The power meter with optical circulator is utilized for back-reflection measurement for comparison to the resistivity measurement.}
\label{fig:overview}
\end{figure}

\section{Sample Fabrication}
The NbTiN film was co-sputtered in a reactive magnetron sputtering system in a mixed atmosphere of N$_2$ and Ar. Film was deposited on commercial c-cut sapphire substrates at room temperature. Two separate source targets with niobium and titanium were installed in the ultrahigh vacuum chamber operating at a base pressure of $1\times10^{-7}$ Torr. The working pressure was kept at 3 mTorr, while Ar and N$_2$ gas flows were 50 and 5 sccm, respectively. The Nb and Ti targets were controlled with a 80 W DC and 240 W RF power sources, respectively. The deposition rates were initially calibrated on thicker films.

The fabrication process involved patterning a meander-shaped nanowire via electron beam lithography in PMMA resist and then transferring the pattern using reactive ion etching (RIE) with CF$_4$ plasma. We fabricated a 7.5~nm thick NbTiN test sample, with sheet resistance $R_s=285~\Omega$. The resulting structure features a 30 µm inner diameter, a 500~nm nanowire width $w$, 500~nm spacing $s$, corresponding to fill factor $f=w/(w+s)$ of 0.5.

The ground plane and meander were wire-bonded with aluminium wire. The transport properties were characterized down to 4~K. The residual resistivity ratio is 0.8 and the critical temperature is 9~K. \begin{figure}[h!]
\centering 
\includegraphics[width=0.55\linewidth]{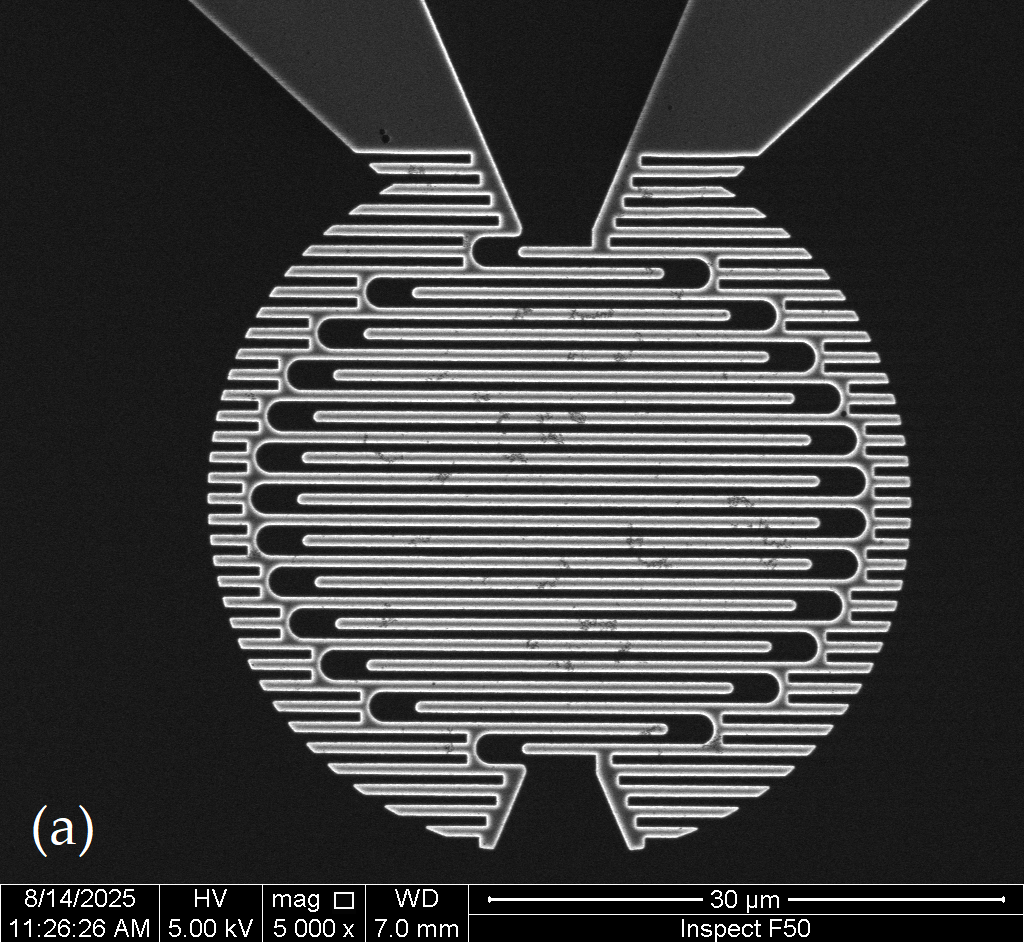}
\includegraphics[width=0.45\linewidth]{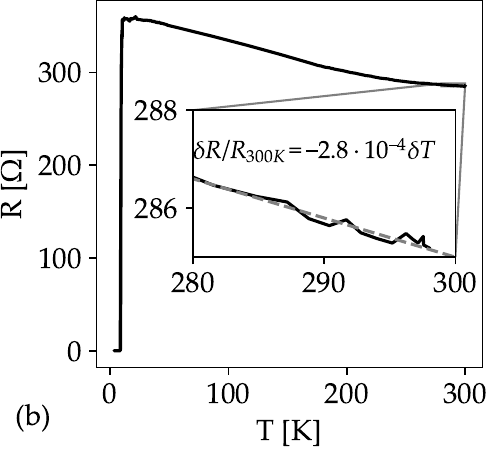}\\
\caption{(a) SEM image of the NbTiN nanowire. (b) Transport measurement of the NbTiN film. The inset shows the detail of the resistivity dependence in the vicinity of room temperature, where the alignment was carried out. The extracted temperature coefficient ratio at room temperature is $\alpha=-2.8\times10^{-4}$.}
\label{fig:SEM}
\end{figure}
The test nanowire, shown in Fig. \ref{fig:SEM}a was fabricated with diameter of 30~$\mu$m and width of 500~nm resulting in a comparable resistance to the common dimensions of $\approx$10~$\mu$m and 100~nm.\cite{Chang21} 
The signal strength is proportional to the change of resistance:
\begin{equation}
    V_{response}\propto\delta R(T) = R(300\mathrm{K})\times\alpha\times(T-\mathrm{300K}),
 \label{eq:resp}
\end{equation}
where $\alpha$ is the temperature coefficient of resistance, given by the residual resistivity ratio.

\section{Optical absorption}
The optical absorption $A$ of a metallic film in the limit of thin films (thickness $d<\lambda$) with wavelength-dependent sheet resistance $R_s(\lambda)$ on a substrate with refractive index $n_{sub}$ is given as follows\cite{Semenov09}:
\begin{equation}
\label{eq:abs}
    A(\lambda) =\frac{4R_s/Z_0}{\big((R_s(\lambda)/Z_0)(n_{sub}+1)+1\big)^2},
\end{equation}

where $Z_0=377\Omega$ is the vacuum impedance. A common approximation is to neglect the wavelength dependence of $R_s$ and simply utilize its DC value estimated by four-probe transport measurement. For example, for our film,  $R_s=285\Omega$ would lead to absorption of $A=31.9\%$. 

\begin{figure}[h!]
    \centering
    \includegraphics[width=0.5\linewidth]{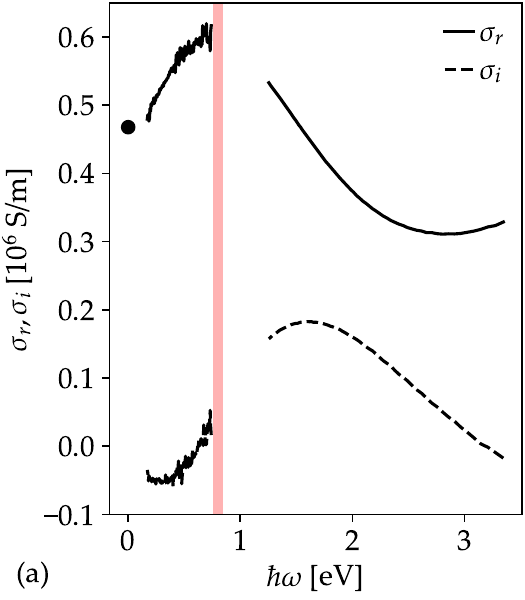}%
    \includegraphics[width=0.5\linewidth]{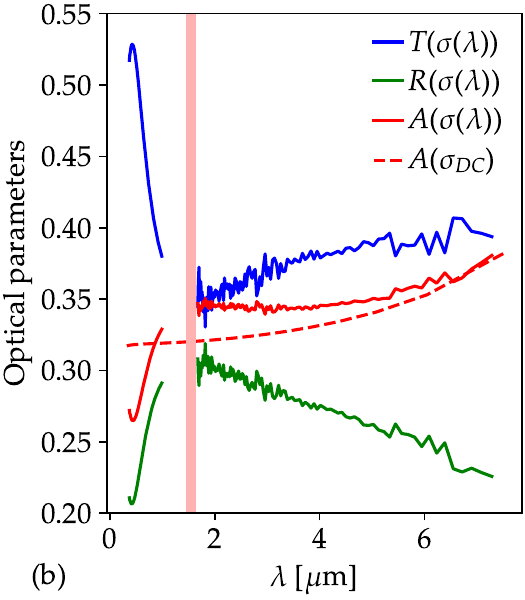}%
    
    \caption{(a) Optical conductivity determined from spectroscopic ellipsometry and DC conductivity from four-point measurement (circle). (b) transmission, reflectivity, and absorbance spectra calculated from conductivity using Fresnel equations at 0° angle of incidence. Red vertical line is guideline for $\lambda=1550$~nm. Red dashed line represents absorption assuming DC sheet resistance of the sample.}
    \label{fig:ellipso}
\end{figure}

However, for strongly disordered films, quantum corrections to the conductivity are present and the DC conductivity $\sigma_{DC}=1/(R_sd)$ is strongly suppressed compared to the optical conductivity at 1550~nm.\cite{Neilinger19,SamoNbN} The optical conductivity of the film $\sigma=\sigma_r+j\sigma_i$ is determined by spectroscopic ellipsometry, shown in Fig. \ref{fig:ellipso}. The DC conductivity is suppressed by 30\% compared to its value at 1550~nm. The corresponding absorption A, reflection R, and transmission spectra T of the thin film, calculated from Fresnel equations, are shown in Fig. \ref{fig:ellipso}. Its value at $\lambda =$ 1550~nm is $A=34.4\%$.

Given that the nanowire width and spacing are significantly smaller than the irradiation wavelength, the effective medium approximation can be used.\cite{Moon06} Therefore, a sufficiently narrow meandering nanowire of film sheet resistance $R_s$ with a fill factor $f$ can be approximated as a thin film with sheet resistance $R_s^{eff}=R_s/f$. 
The feasibility of the effective medium approximation for common SNSPD designs was also verified numerically in Ref. \onlinecite{Baranek24}.

The resulting absorption of the fabricated nanowire is $A^{NW}=26.3\%$ for the conductivity at 1550~nm. In the case of 500~nm wide separation between nanowires, the absorption may decrease due to the light leakage. Moreover, polarization dependence is stronger for sub-100~nm nanowires.

The temperature distribution in thin films can be determined either through analytical methods (see, for instance, Ref. \onlinecite{Tadeau04}) or through numerical simulations, which allow for the entire meander to be modeled. The frequency and displacement dependence of the photothermal response is governed by thermal properties of substrate, thin film, and their interfacial thermal resistance (Kapitza resistance, see, e.g. Ref. \onlinecite{Nguyen19}).

\section{Apparatus}
\begin{figure}[h!]
\centering 
\includegraphics[width=0.6\linewidth]{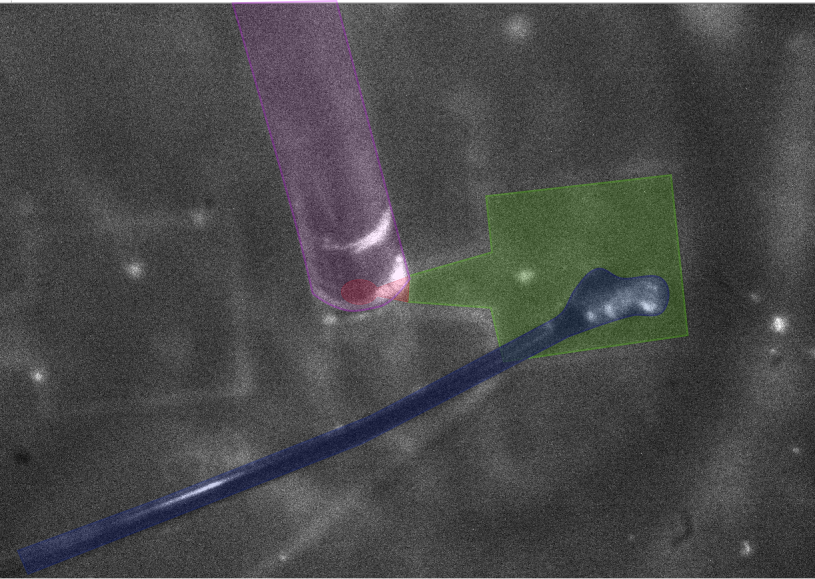}

\caption{Pseudocolored photo of fiber (purple) suspended over meander (red). Bonding pad of the meander (green) is aluminum wire-bonded (blue). Grounding wire-bonds are not visible in the photo.}
\label{fig:reallife}
\end{figure}

The wirebonded nanowire was placed on a PCB, which was attached to the motorized XYZ nano-positioning stage (Attocube ECSxy5050, ECSz5050), as shown in Fig. \ref{fig:reallife}. First, the fiber was positioned at starting position utilizing manual micro-positioners. Then, the scan was performed with nano-positioners with 50~nm repeatability.

The illuminating laser source (HP 8168E, $\lambda=$1550~nm) was AC modulated and the photothermal response was measured by SRS SR830 Lock-in amplifier locked to the laser modulation frequency $f_{mod}$ with the AC-coupled FEMTO DLPVA-100-B-D preamplifier. To prevent damage to the lock-in/preamplifier, additional DC blocks were utilized.

The DC bias was realized as a balanced differential voltage through a pair of 12V A27 batteries, to eliminate ground loops, noise from the current source, and interference caused by high voltage pulses from piezo actuators in the nano-positioner. The small size of the batteries allowed them to be placed in close proximity to the nanowire. The ground was then connected to the apparatus grounding point. All wires were twisted pairs with grounded shielding. The sample on a PCB was placed on top of the grounded baseplate.

\begin{figure}[h]
\centering 
\includegraphics[width=0.6\linewidth]{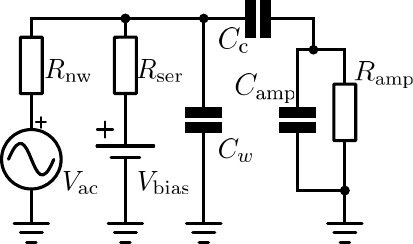}%
\caption{Equivalent scheme for the readout circuit.}
\label{fig:Replace}
\end{figure}

To maximize the output signal at given bias voltage $V_{bias}$, the serial resistance $R_{ser}$ should be impedance-matched to the nanowire resistance $R_{nw}$ (e.g. 400~k$\Omega$). This biasing circuit provided 30~$\mu$A of bias current/12~V of bias voltage to the nanowire, resulting in 30~$\mu$V response.
The equivalent scheme for the readout circuit is presented in Fig. \ref{fig:Replace}, where $C_w$ is the cable capacitance, $C_c$ is the AC coupling capacitance, and $R_{amp}$ and $C_{amp}$ are input resistance and capacitance of the first amplification stage. The AC voltage generated at nanowire is given as $V_{AC}=\delta R\times V_{bias}/(R_{nw}+R_{ser})$, where $\delta R$ is given by eq. \ref{eq:resp}.

\section{Results}

\begin{figure}[h!]
\centering 
\includegraphics[width=\linewidth]{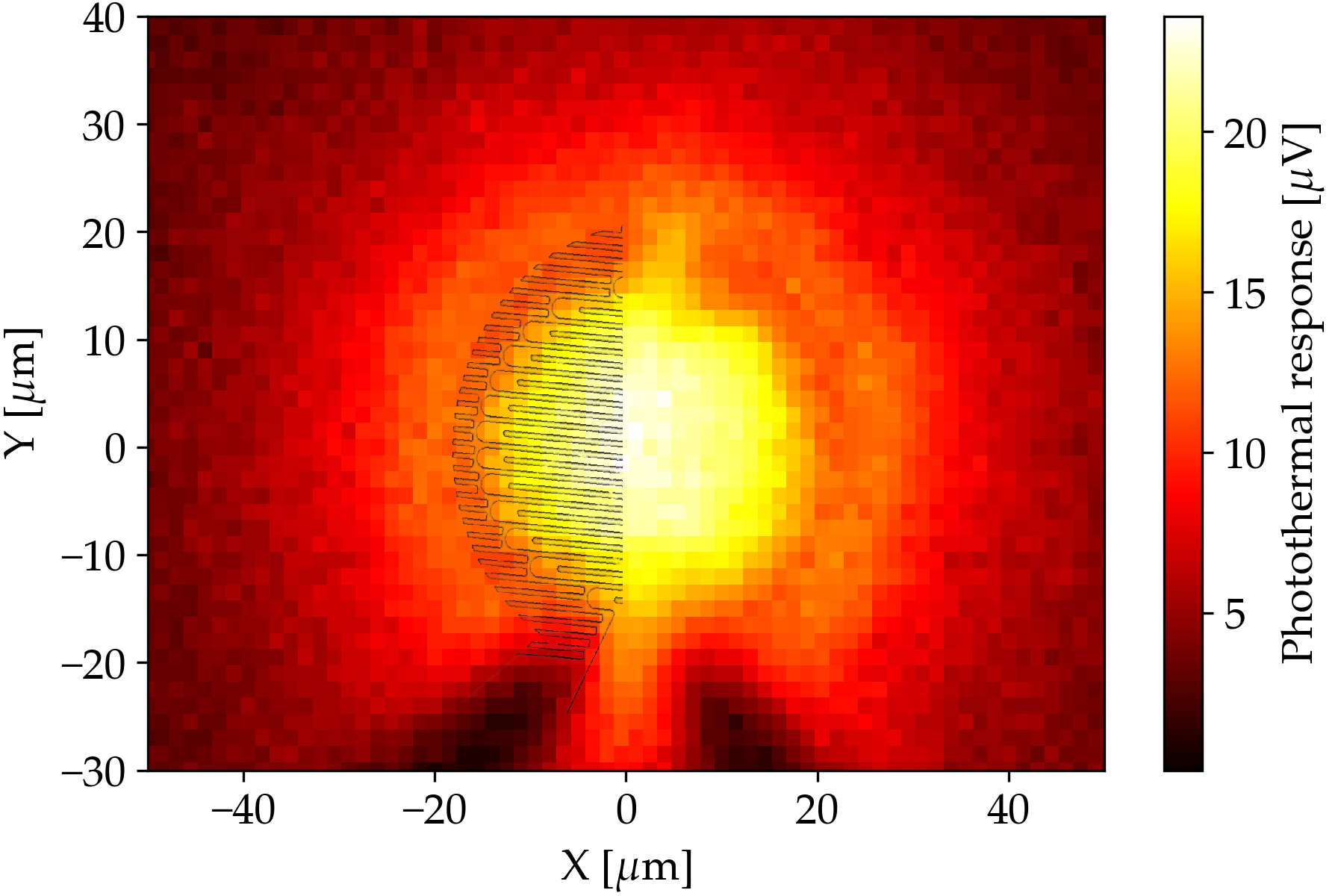}

\caption{XY scan of photothermal response at lowest height, with 1.5~$\mu$m step. To visualize the detection sensitivity, we superimposed SEM image of the test nanowire from Fig. \ref{fig:SEM}a with enhanced contrast.}
\label{fig:xy_fine}
\end{figure}

\begin{figure}[h!]
\centering 
\includegraphics[width=0.6\linewidth]{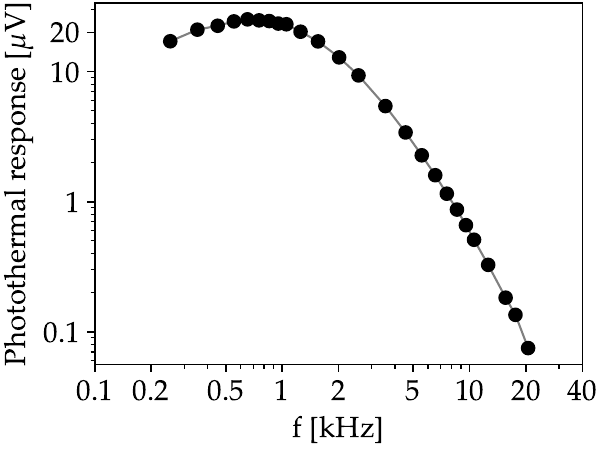}%
\caption{Dependence of photothermal response on laser modulation frequency $f_{mod}$. Low frequency response is limited by the AC coupling of the circuit, while high frequency response is limited by high impedance and cable capacitance.}
\label{fig:response}
\end{figure}

The XY-scan of the photothermal response at lowest height with 1.5~$\mu$m step is shown in Fig. \ref{fig:xy_fine}. The maximum response amplitude, corresponding to the perfect alignment of the fiber and the meander-shaped nanowire, is 23~$\mu V_{RMS}$. When the fiber is far from the center of the nanowire, the temperature and thus the resistance of the nanowire are not modulated, and the signal amplitude drops below 1~$\mu$V, corresponding to SNR of approximately 20.  
Due to the high output impedance and the low signal amplitude, the resulting signal current is $\le$100~pA, thus, the choice of modulation frequency is limited by the capacitance of the wires ($C_w$). This can be seen in Fig. \ref{fig:response}, where a high frequency cutoff is clearly visible at frequencies of $\approx$1~kHz. Therefore, the XY scan was performed at $f_{mod}=$~871~Hz.

\begin{figure}[h!]
\centering 
\includegraphics[width=0.5\linewidth]{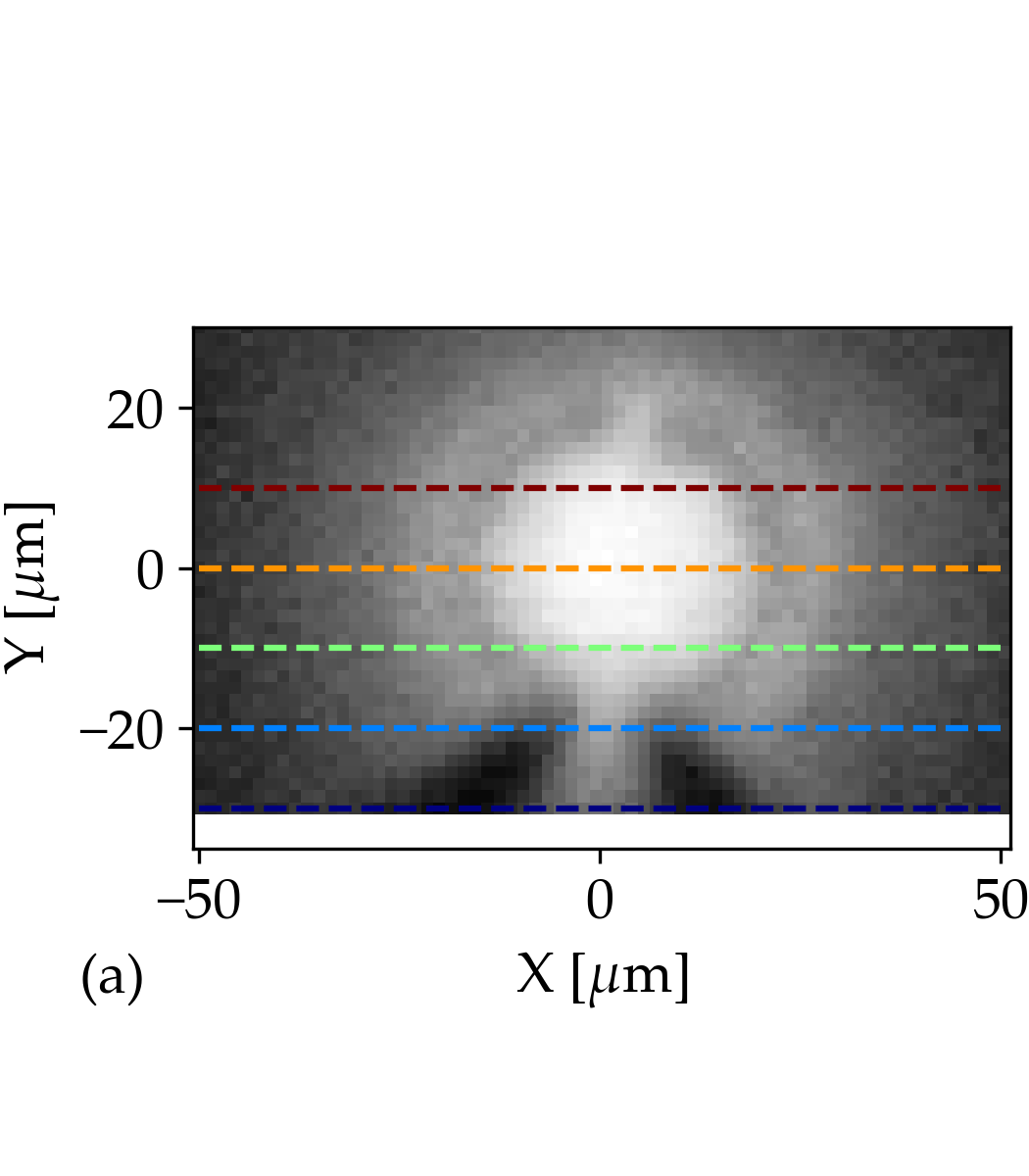}
\includegraphics[width=0.5\linewidth]{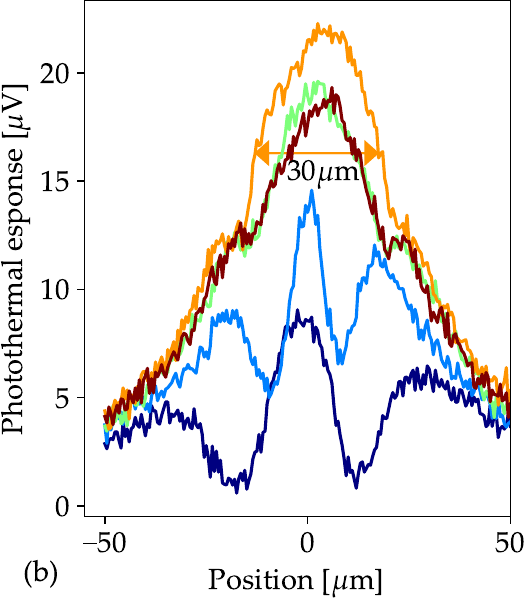}%

\caption{(a) XY scan of photothermal response. Horizontal lines show the cross-sections. (b) Cross-sections measured with 500~nm steps.}
\label{fig:xy_cross}
\end{figure}

The cross-sections measured with 500~nm steps in Fig.\ref{fig:xy_cross} show that the resolution is limited by the beam width and by the noise of the measurement, rather than by the precision of the nano-positioning. 

\begin{figure}[h!]
\centering 
\includegraphics[width=\linewidth]{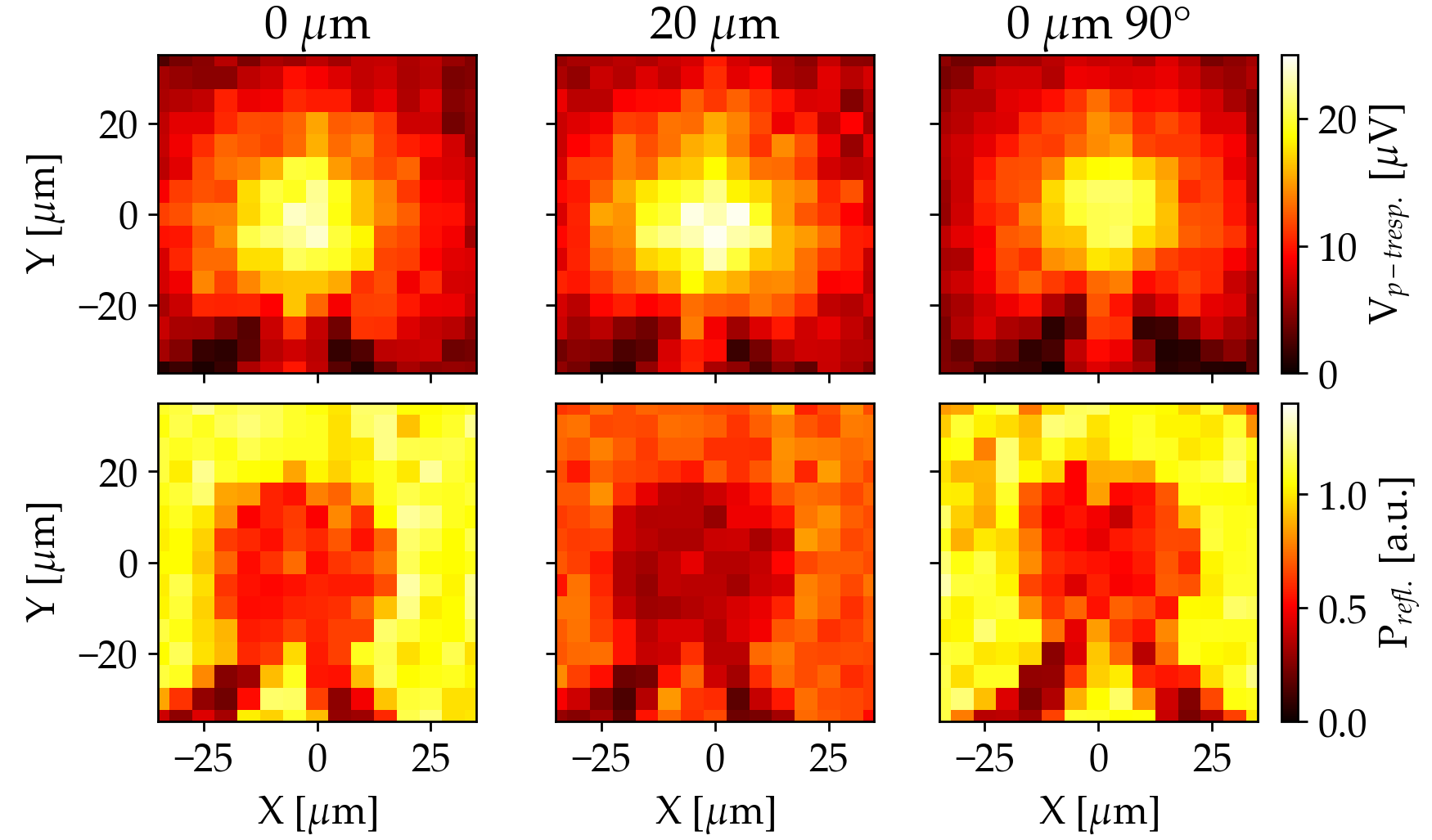}%
\caption{Photothermal response of the SNSPD (top row) and back-reflection measurement (bottom row) at different heights (left and center) and for sample rotated by 90° (right) to test the polarization sensitivity.}
\label{fig:ref_vs_abs}
\end{figure}

To compare the performance of the photothermal response, we performed back-reflection optical measurement, utilizing an optical in-fiber circulator and Thorlabs S154C photodiode in PM100D power meter, shown in Fig. \ref{fig:overview}.
The scans are shown in Fig. \ref{fig:ref_vs_abs}. The polarization dependence of a 500~nm wide nanowire ($w\approx\lambda/3$) differs significantly from the sub-100~nm ($w\approx\lambda/15$) state-of-the-art SNSPDs.

\begin{figure}[h!]
\centering 
\includegraphics[width=\linewidth]{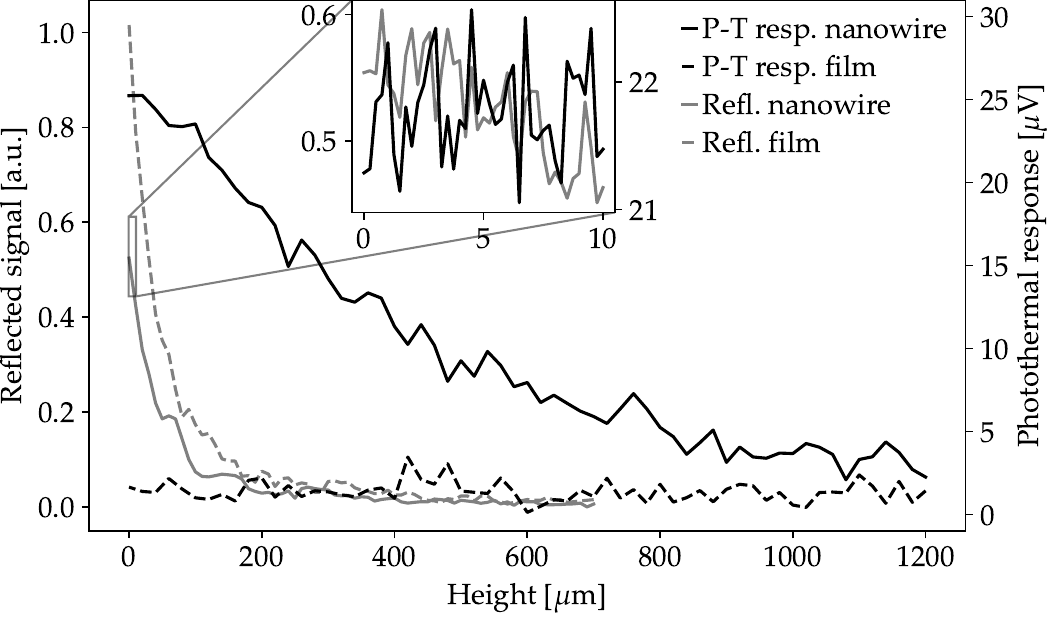}%
\caption{Comparison of the height scans of the photothermal and back-reflected method. The inset shows the detail of the height scans in the vicinity of the surface ($<$10~$\mu$m) with 250~nm steps.}
\label{fig:zsweep}
\end{figure}

The comparison of the height sweep scans (presented in Fig. \ref{fig:zsweep}) reveals a significantly stronger height dependence in the back-reflection measurement, as it requires coupling of the diverging light back into the 10~$\mu$m fiber core.

\section{Conclusion}
We have proposed and demonstrated the feasibility of a simple photothermal alignment technique for SNSPDs that exploits the temperature dependence of the nanowire resistance to generate a signal detectable by a lock-in upon local heating by a modulated optical beam irradiating the active area. Using a laser with modulation intensity of $\approx$~400~$\mu$W, we obtained a peak response of 23~$\mu$V$_{\mathrm{RMS}}$ with signal-to-noise ratio  $\mathrm{SNR}\approx$20 and spatial resolution in the range of $\approx\mu$m, with fine height scans resolved to 250~nm and XY scans to 500~nm.

The presented method requires 3D alignment means similar to optical methods, such as back-reflection measurement, with the advantage of lower sensitivity to the fiber height and incident angle misalignment. Thus, this photothermal alignment is an accessible method that is fabrication-compatible with the SNSPD production and can substitute existing fiber-to-chip alignment methods or complement them to improve reproducibility and throughput in SNSPD finalization. Moreover, the presented method can be used for in-situ coupling efficiency evaluation of the SNSPD detector during cooldown and could be utilized in other fields of physics, such as, for example, in the study of heat transfer in nanostructures,\cite{Sidorova24,Romeo25} in thin films in general,\cite{GrosseKockert22,Dames13} or as an additional independent measurement channel for time-domain thermoreflectance measurements \cite{Yuan22,Mohan25} for sufficiently conductive samples. Although beyond the scope of this paper, we are currently investigating the feasibility of using this method to extract the heat capacity and thermal conductivity of the thin films.

\section*{Funding}
This work was supported by the project skQCI (101091548), funded by the European Union (DIGITAL) and the Recovery and Resilience Plan of the Slovak Republic.

\section*{Conflict of Interests}
The authors have no relevant financial or non-financial interests to disclose.

\section*{Author Contributions}
Conceptualization - PN (lead), MB (supporting); Data Curation - MB (lead); Formal Analysis - MB (lead), PN (supporting); Funding Acquisition - MG (lead), PN (supporting); Methodology - MB (equal), PN (equal), MG (supporting); Project Administration - MG (lead), PN (supporting); Resources - MB (equal), PN (equal),  DL (equal), T\v{S} (equal), J\v{S} (equal), IV (equal), \v{S}H (equal), MG (supporting); Software - MB (equal), DL (equal); Supervision - PN (lead); Validation - MB (lead), DL (supporting), PN (supporting); Visualization - MB (lead); Writing/Original Draft Preparation - MB (equal), PN (equal), J\v{S} (supporting), T\v{S} (supporting); Review and Editing - all authors (equal).


\section*{Data Availability Statement}
The data that support the findings of
this study are available from the
corresponding author upon reasonable
request.

\nocite{*}
\bibliography{aipsamp}

\end{document}